# Non-Adhesive Transfer Process of Carbon Nanotube Forests onto Flexible Kapton Substrates

*Terry Lukov, Robinson Smith, Ashley R. Chester*

*Rio Salado College, Tempe, AZ, USA*

***Abstract:*** *Prevalence of electronic gadgets has been on the rise and among several considerations that are important for furthering frontiers of utility of new gadgets is mechanical flexibility of electronic systems. Flexible electronics require functional materials to be held by flexible substrates that can be bent to small radii of curvature. In addition, the functionality of the active material has to remain intact during the process of bending. There have been challenges in the transfer process of functional materials onto flexible substrates as well as challenges in maintaining the functionality of these materials during the bending process. Here we demonstrate transfer of carbon nanotubes, a highly studied electronic material, onto flexible substrates, mainly kapton. We study the optimum conditions of adhesion, pressure and temperature required to obtain the strongest transferred layer. We further study the limits on the radius of curvature that the substrate can be bent to.*

As the Moore's law struggles to extend the limits of miniaturization of electronics, there have been several innovations to uphold the continued addition of utilities and miniaturization of electronics. The effort to pack more utility into consumer electronics involves building mechanically flexible electronic devices that have several application ranging from flexible displays to smart watches. The research into manufacture of flexible electronics primarily involves four aspects: (1) choice of a functional electronic material, (2) choice of a flexible substrate, (3) transfer process involving transfer of the active material onto the flexible substrate, and (4) achieving small radii of curvature without compromising on the functionality of the system.

Here we address each of the preceding four topics. We chose to use carbon nanotubes (CNTs) as the active electronic material, owing to their excellent applications in sensing, information storage, energy harvesting, batteries, actuation, etc. We then chose kapton as the flexible substrate due to the great interests it has generated and its compatibility with several aspects of semiconductor processing. We then describe the transfer process and explore optimal conditions for transfer including temperature, pressure, substrate and growth conditions. We then experiment on the extent of bending the substrate with the transferred CNTs can withstand before they show deviation in their electrical behavior. Table 1 provides a summary of comparisons between Si based electronic devices and CNT based composite materials. It is clear that despite the non-traditional fabrication procedures involved in CNT based electronics, they provide far more flexibility in terms of range of utilities and mechanical flexibility of substrates.

|  | Silicon based | CNT based |
| --- | --- | --- |
| Fabrication process | Traditional | Non-traditional – involves chemical steps too |
| Substrate | Si-SiO2 | polyvinylidenefluoride (PVDF) |
| Stretchable? | No<br>very brittle (not possible to make a sensor of in-plane stress) | Yes<br>140% - not brittle (possible to make a sensor of in-plane stress) |
| Integration with Kapton | Integration with Kapton is not easy, since we cannot grow silicon on Kapton | Since Base is also a polymer, integration might be possible when curing is done over Kapton |
| Temperature | Can stand process temperatures of more than 700 C | Can stand temperatures of 200C or 400F |

**Table 1:** Comparison of Si based electronic devices and CNT based composite materials

Growth of CNTs was done using a previously described process of carbonation of a thin film catalyst. A lightly boron doped Si substrate of thickness 500 μm was chosen, which was made to undergo a RCA cleaning process in order to remove contaminants, ionic impurities, carbonaceous impurities and oxide-based residue. Following this, it was baked in vacuum to remove moisture. 5 nm of Al and 4.5 nm of Fe was deposited uniformly on the entire substrate using e-beam induced evaporation. The substrates were then cleaned with heated acetone and iso-propyl alcohol for 10 minutes and dried with nitrogen gas. We used a tube furnace of diameter ~1 inch into which the substrates were loaded. The process first involved cleansing the system with Ar at room temperature for 30 minutes at a partial pressure of 1000 mTorr. Then the temperature was ramped up to 760 C over 15 minutes while maintaining the Ar environment. Once the temperature stabilized, the growth process initiated. Ethylene gas was flowed at a partial pressure of 450 mTorr at 760 C along with hydrogen flow at a partial pressure of 150 mTorr. This lasted 6 minutes. Following this step, the environment changed to Ar flow at a partial pressure of 1000 mTorr for 10 minutes without changing the temperature. This was followed by a ramp down of temperature to room temperature over 3 hours. Figure 1a is a scanning transmission micrograph of a forest of vertically aligned CNTs that were about 160 μm in height. Figure 1b is an optical photograph of a number of substrate pieces with only one of them coated with the CNT forest, apparent from the very low reflectivity of light.

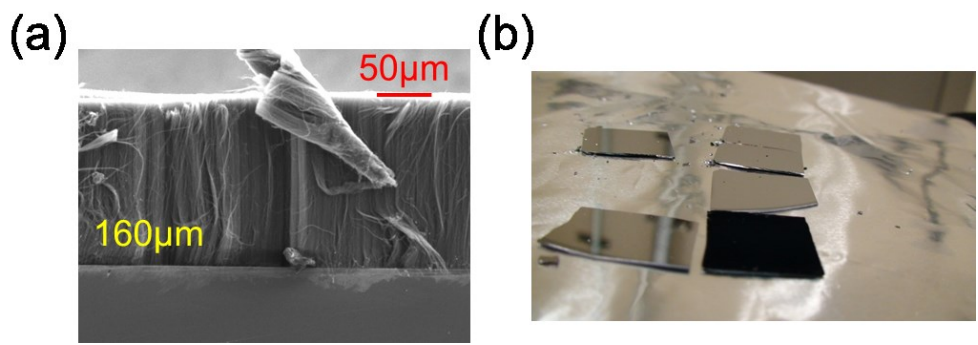

**Figure 1:** (a) Cross sectional SEM of the CNT forest grown. (b) Photograph of several substrates with only one with the CNT forest, identified by the black color and low reflectivity.

We then attempted to transfer the CNT layer onto kapton. This was achieved by using an adhesive kapton tape that was commercially available. The kapton tape was lightly pressed against the CNT surface and then peeled off by hand. Figure 2a is a photograph of the CNt layer transferred onto the kapton tape. In order to test the strength of adhesion of the CNT forest onto the kapton tape, we used an ultra-sonication test. Here, we immersed the tape with CNT forest into a water bath and subjected it to ultrasonication for several minutes. After a certain interval of time, the CNT forest was inspected to note peel-off or damages. Figure 2b-2d show photographs of the same CNT forest after subjecting to ultrasonication for different extents of time, using the process described here. It is fairly apparent that the CNT forest did not show visible signs of extensive damage or peel-off after 30 minutes of ultrasonication. We followed up this experiment by examining if the CNT forest could be transferred to other substrates. For this, we chose a 3M fabric (commercially available) tape with adhesive on one side, using a transfer process similar to the one described in this paragraph. The resulting transfer is shown in Figure 2e. This was followed by an attempt to transfer the CNT forest onto kapton tapes without adhesives. To improve adhesion, we used high temperatures and pressures. As an example, the process utilizing 100 C and 200 kPa of temperature and pressure, respectively, produced the transfer shown in Figure 2f.

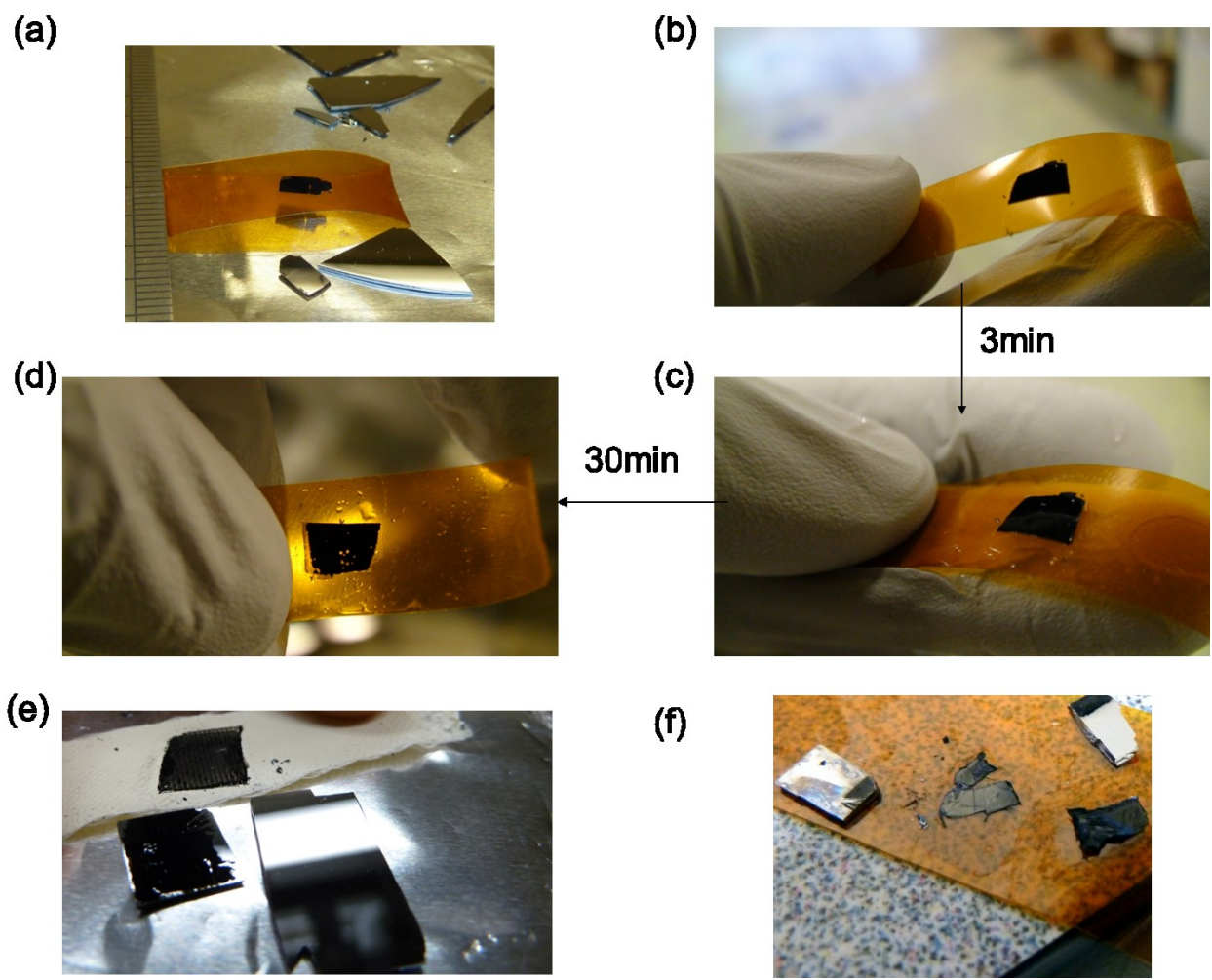

**Figure 2:** (a) Photograph of the CNt layer transferred onto the kapton tape with adhesive. (b) Photographs of the same CNT forest after subjecting to ultrasonication for different extents of time, using the process described here. (e) CNTs transferred onto a 3M fabric (commercially available) tape with adhesive on one side, using a transfer process similar to the one described here. (f) Transfer process without using adhesives.

We followed up with a systematic study of the adhesion strength of the CNT forests onto the substrates. For this, we subject the forests of transferred CNTs to ultrasonication, as described in the previous paragraph, for different durations of time. After each time interval, we inspect the surface coverage relative to the initial coverage. In the course of ultrasonication, due to damages and peel-off, a part of the forest tends to fall off the flexible substrate. This was used as a measure of the adhesion strength. The first experiment involved transfer of three different forests of CNTs at a temperature of 100 C, using three different pressures, onto kapton substrates without using an adhesive. The results are plotted in Figure 3a. The curves show that with increasing time of ultrasonication, there is considerable damage to the CNT forest, especially after several tens of minutes of ultrasonication. Another interesting aspect of the plot is the non-monotonic dependence of the adhesion strength on the pressure used for transfer. The forest transferred using 200 kPa of pressure appears far more stable than those transferred using either 100 kPa or 300 kPa. This implies an optimum pressure that is required to transfer CNTs onto flexible substrates. With too little pressure, there may not be sufficient localized attractive forces, including forces due to locally created electrostatic charges and van der Waal's forces. On the other hand, too high a pressure also means bending of the CNT forests such that they lose their mechanical stability. We also showed that transfer onto a flexible fabric as described above yielded similar results compared to kapton (Figure 3a). This also shows that the transfer process can be generalized to many substrates.

We then explored the role of temperature in the transfer of CNT forests onto kapton substrate. We repeated the experiment of ultrasonication described above on different transferred CNT forests. All forests were transferred using a pressure of 100 kPa, but with different temperatures, namely 40 C, 100 C and 200 C. We found that the forests transferred using 200 C disintegrate very quickly (Figure 3b). The trend in temperature was found to be similar to the trend in pressure applied, that is, there was an optimal temperature of transfer, above or below which the transfer was not as optimal. The explanation for this could be similar to the explanation offered to justify the effect of pressure. A high temperature might be required to activate the localized van der Waal's forces to be accessed (similar to localized ultrasonication to bond wires). A very high temperature might burn the interface, leaving it brittle. A very low temperature might not access the van der Waal's forces. We found that 100 C was the best among the temperatures experimented with.

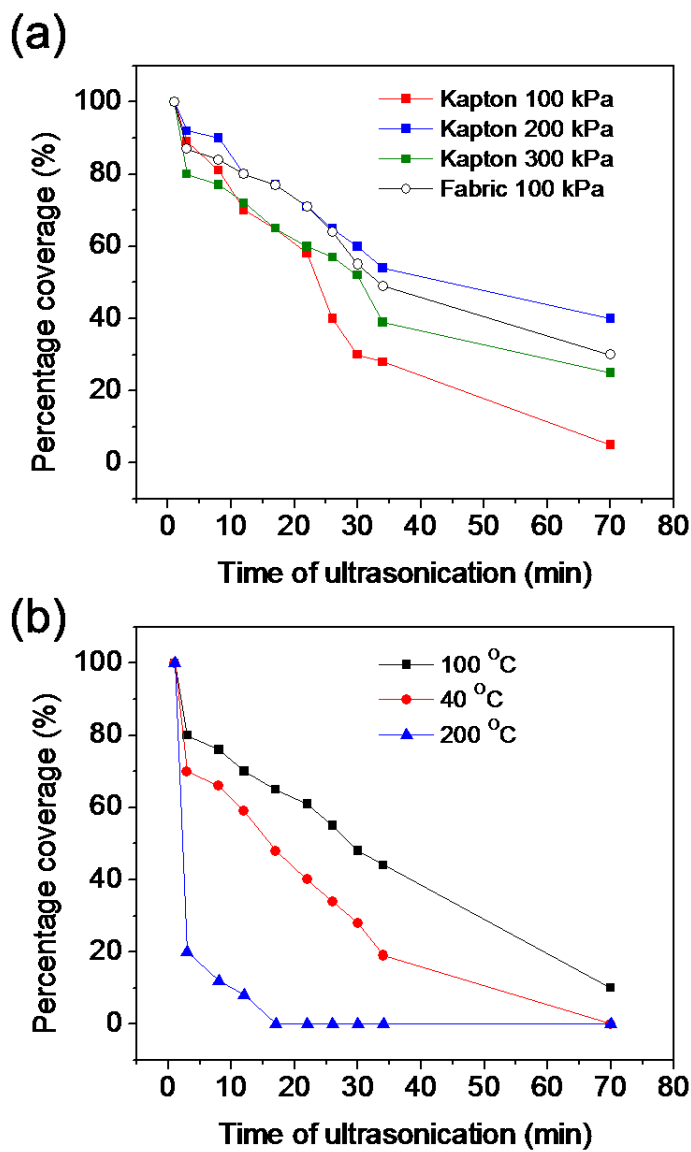

**Figure 3:** (a) Ultrasonication test of transfer of three different forests of CNTs at a temperature of 100 C, using three different pressures, onto kapton substrates without using an adhesive. Also shown is the results from the fabric used. (b) The experiment in (a) repeated with a oncstant pressure (100 kPa) and variable temperature, as noted.

We then studied the failure, damage, or peel-off, using a more microscopic analysis, utilizing a scanning electron microscope (SEM). We use two quantities: L is the length of CNTs and H is the height of the CNT forest. So H can be less than or equal to L, depending on how much the CNTs are bent. A clear correlation emerges between mechanical microstructure and adhesion. Four different regions could be identified corresponding to different failure modes, directly related to the L/H ratio. The relationship is represented in Figure 4c. When L/H<1, the failure occurs in the delta layer. When the ratio is 1<L/H<3, there are two failure mechanisms: cohesive failure on the entire bond and (b) failure on the support layers. When L/H>3, there is failure of the adhesion. Figure 4a is an example of a fracture propagating in a forest of CNTs that was transferred onto kapton tape with no adhesive and Figure 4b is an example of a fracture propagating in a forest of CNTs that was transferred onto kapton tape with adhesive.

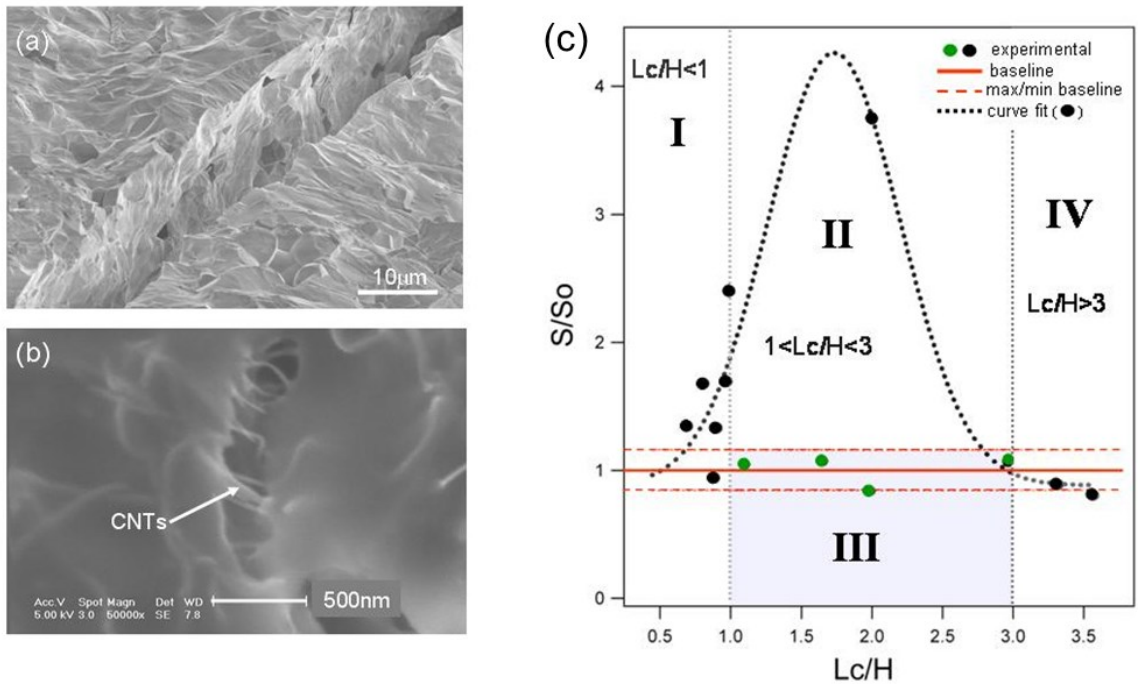

**Figure 4:** (a) SEM image of a fracture propagating in a CNT forest with no adhesive, and (b) SEM image of a fracture propagating a CNT forest on adhesive. (c) Plot of different regions of failures corresponding to the L/H ratios.

Further, we also measure the adhesion strength using a third technique, namely electrical conductivity or resistance of the film. This is a macroscopic measurement, but provides information on the microstructure because the currents flowing within the film/forest is a function of the percolating pathways available for conduction. In this experiment, instead of using ultrasonication, we bent the substrate at different radii of curvature. This is a direct measure of the utility of such a substrate in flexible electronics. The two CNt forests considered were that transferred onto kapton with no adhesive and that transferred onto kapton with adhesive. The resistance between two opposite edges of the CNT forest appears to increase with a decreasing radius of curvature (Figure 5). This can be easily understood intuitively. As the substrate is bent further, there are cracks appearing in the CNT forest. This breaks the conduction pathways, thereby increasing the resistance of the forests. It is observed that the forest without adhesive undergoes an increase in resistance slightly more pronounced than that with an adhesive. This is not a significant difference and the difference is not as much in the radius of curvature at which this departure appears. So it is possible to conclude that transferring CNTs onto kapton without using adhesives appears to be as promising as transferring CNTs onto kapton with adhesive.

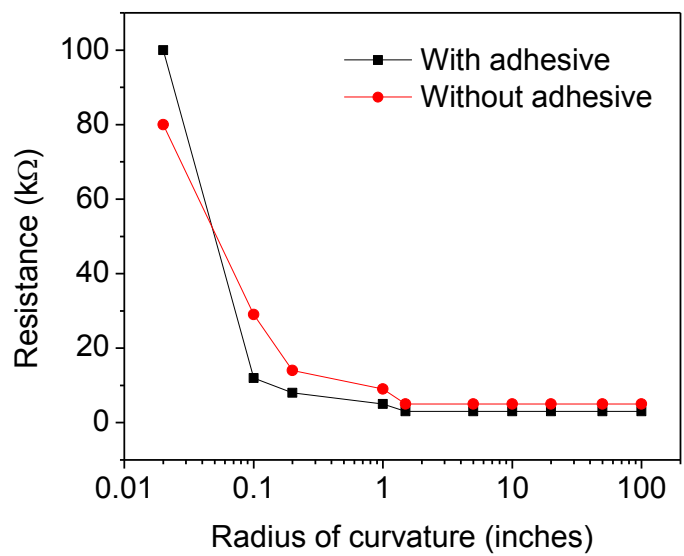

**Figure 5:** Dependence of resistance of the CNT forest on radius of curvature for CNTs transferred onto kapton with and without adhesive.

In conclusion, we performed several experiments of transferring CNTs onto different flexible substrates, with a focus on kapton. We examined the endurance of the CNT forests to mechanical stresses in the form of ultrasonication and also studied the failure mechanisms using SEM. We found that there is an optimal pressure and temperature that allow for the best transfer of CNTs onto kapton without using adhesives. We also showed that the electrical performance of CNTs transferred onto kapton without using adhesives can be as good as that transferred onto kapton with adhesives.

These results provide a pathway for transferring a promising electronic material, namely CNTs, onto flexible substrates like kapton using an easy process that will pave the way for future flexible electronics.